\documentclass[sn-mathphys,Numbered,iicol]{sn-jnl}% Math and Physical Sciences Reference Style
\usepackage{changes}
\usepackage{graphicx}%
\usepackage{multirow}%
\usepackage{amsmath,amssymb,amsfonts}%
\usepackage{amsthm}%
\usepackage{mathrsfs}%
\usepackage[title]{appendix}%
\usepackage{xcolor}%
\usepackage{textcomp}%
\usepackage{manyfoot}%
\usepackage{booktabs}%
\usepackage{algorithm}%
\usepackage{algorithmicx}%
\usepackage{algpseudocode}%
\usepackage{listings}%
\usepackage{subfigure}%
\usepackage{graphicx}

\theoremstyle{thmstyleone}%
\theoremstyle{thmstyletwo}%

\theoremstyle{thmstylethree}%

\begin{document}

\title{A novel LSTM music generator based on the fractional time-frequency feature extraction}


\author[1]{\fnm{Li} \sur{Ya}}\email{liya@hainnu.edu.cn}

\author[2]{\fnm{Chen}\sur{Wei}}\email{chenwei@hainanu.edu.cn}
\equalcont{These authors contributed equally to this work.}

\author*[3]{\fnm{Li}\sur{Xiulai}}\email{lixiulai01@hainanu.edu.cn}
\equalcont{These authors contributed equally to this work.}

\author[1]{\fnm{Yu}\sur{Lei}}\email{yulei12@hainnu.edu.cn}
\equalcont{These authors contributed equally to this work.}

\author[1]{\fnm{Deng}\sur{Xinyi}}\email{dengxinyi@hainnu.edu.cn}
\equalcont{These authors contributed equally to this work.}

\author[3]{\fnm{Chen}\sur{Chaofan}}\email{cfchen@acrsea.com}
\equalcont{These authors contributed equally to this work.}

\affil[1]{\orgdiv{College of Music},\orgname{Hainan Normal University}, \orgaddress{\street{Longkunnan}, \city{Haikou}, \postcode{570100}, \state{Hainan}, \country{China}}}

\affil[2]{\orgdiv{College of Foreign Language},\orgname{Hainan Normal University}, \orgaddress{\street{Longkunnan}, \city{Haikou}, \postcode{570100}, \state{Hainan}, \country{China}}}

\affil[3]{\orgdiv{R\&D}, \orgname{Hainan Hairui Zhong Chuang Technol Co. Ltd.}, \orgaddress{\street{People}, \city{Haikou}, \postcode{570228}, \state{Hainan}, \country{China}}}


\abstract{In this paper, we propose a novel approach for generating music based on an artificial intelligence (AI) system. We analyze the features of music and use them to fit and predict the music. The fractional Fourier transform (FrFT) and the long short-term memory (LSTM) network are the foundations of our method. The FrFT method is used to extract the spectral features of a music piece, where the music signal is expressed on the time and frequency domains. The LSTM network is used to generate new music based on the extracted features, where we predict the music according to the hidden layer features and real-time inputs using GiantMIDI-Piano dataset. The results of our experiments show that our proposed system is capable of generating high-quality music that is comparable to human-generated music.}



\maketitle

\section{Introduction}\label{sec1}

Nowadays, music can be divided into many categories, such as classical, pop, jazz, electronic, etc.,\cite{2015Statistical} Artificial Intelligence (AI) composition this novel creative means not only meets the needs of different musical styles but also from different dimensions of music innovation and adjustment,\cite{2021Music} the most intuitive of which is tone, as a feature closely related to the signal frequency, the tone is often directly related to the overall feeling of a piece of music. Secondly, the duration of each syllable, and even the adjustment of rhythm, can be automatically generated by AI, to achieve more accurate and higher music quality. It can be seen that AI composition has great potential in the field of artificial intelligence and can promote the development of artificial intelligence. In addition, from the perspective of music education, AI composition can create new musical styles and expressions, thereby expanding the expression and style of music. It also improves the efficiency of music education and helps students better understand music theory and practical skills.

In recent years, AI has made major strides and found use in many fields, including the creation of music. As AI opens up new avenues for musical expression and innovation, this field of study is gaining more and more attention\cite{doush2020automatic}. \cite{briot2020deep} points out the challenges and directions for music generation using deep learning. It indicates that applying deep learning directly to generate content quickly reaches its limits, as the generated content tends to mimic the training set without showing true creativity. In \cite{jiang2020rl}, a deep reinforcement learning algorithm for online accompaniment generation is presented, with the possibility for interactive duet improvisation between humans and machines in real time. The model is developed using training data that is both monophonic and polyphony. The reward model's efficiency is what makes this program work. The network is modified in \cite{huang2021pop}, where pop piano music is generated by deep learning methods. The Transformer-XL model is a condensed version of the original, lacking the mask section. The generated audio can be produced with greater integrity thanks to the simplified Transformer-XL. Transformer is also used in \cite{li2023transformer}, The authors suggest using a transformer-based sequence-to-sequence model with a pre-trained encoder and decoder to create chord progressions for tunes. A decoder utilizes this information to generate chords asynchronously and then produces chord progressions. A pre-trained encoder extracts contextual information from melodies. The suggested approach takes into account the harmony between chord progressions and melodies while addressing length restriction problems.

\cite{briot2020deep} examines the constraints associated with the utilisation of deep learning techniques in the domain of music creation. The concerns are to the absence of oversight in managing generated content, the inclination of models to imitate the provided training data, and the independent functioning of deep learning models. In \cite{briot2021artificial}, a tutorial is proposed that focuses on the application of deep learning techniques in music generation. It examines the initial advancements made in the late 1980s pertaining to the utilisation of artificial neural networks in the domain of music generation, and evaluates their significance in relation to contemporary methodologies. In \cite{amaral2022adaptive}, a novel framework was devised to facilitate the synthesis of video game music utilising the Transformer deep learning model in the field of architecture. The present approach facilitates the customization of music generation for players based on their individual tastes. This is achieved by the selection of a training corpus that aligns with their desired musical style. The software produces a variety of musical layers by employing industry-standard layering techniques commonly utilised in the composition of video game soundtrack. The primary significance of \cite{piletskiy2020development} resides in its examination of Singular Value Decomposition (SVD) as a proficient approach for constructing recommender systems, particularly in the context of music recommendations. The SVD method is recognised as a commonly employed methodology for reducing the dimensionality of data and enhancing the precision of suggestions through the elimination of latent components. This methodology facilitates the provision of tailored and precise suggestions, thereby tackling the issue of delivering customised content to individuals. \cite{he2022improved} introduces a revolutionary music recommendation method that utilises the attention mechanism of a deep neural network. The proposed approach entails enhancing the preprocessing of MFCC audio data and integrating it with user-specific portrait attributes. An Artificial Intelligence Network (AIN) Recurrent Neural Network (RNN) is employed to generate a curated selection of recommendations by utilising the user's historical listening data. The utilisation of this methodology yields enhanced precision in recommendations as compared to conventional collaborative filtering methods. In \cite{jeong2022multi}, the authors present a unique methodology referred to as the Multi-Objective Generative Deep network-based Estimation of Distribution Algorithm (MODEDA). The present approach is centred on the resolution of challenges related to high dimensionality in large-scale multi-optimization problems (LSMOP) that arise in the context of evolutionary algorithm music creation. In order to address these problems, we suggest employing dimensionality reduction in the decision space and utilising a distinct solution search methodology that optimises within the converted space while upholding consistency with the Pareto sets of the original problem. The algorithm's optimisation performance and computing efficiency in the context of LSMOP are demonstrated by experimental results obtained from knapsack problems and music composition trials.

LSTM is another well-used method to generate music. The ML model is based on LSTMs which takes in previous notes and predicts the next set of notes based on a midi format. For the rule-based method, chord progression rules and binary rhythm pattern theory are applied in \cite{wiriyachaiporn2018algorithmic} and both algorithms are used to generate music. LSTM neural networks by themselves do not produce natural music that conforms to music theory. LSTM combined with grammar inspired by music theory\cite{sun2018composing}. The main principles of music theory are encoded as grammatical argument (GA) filters on training data so that machines can be trained to generate music that inherits the naturalness of human creations from the original dataset while adhering to the rules of music theory. \cite{xiao2016music} proposes a method of using the LSTM Unit model to format music files, extract characters and generate music. And then constructing an automatic music generation system through formalization processing and neural network supervised learning. Paper \cite{toh2021generation} focuses on the music elements, such as pitch, time, and velocity, which are extracted from MIDI tracks and encoded with piano-roll data representation. And a Deep Convolutional Generative Adversarial Network (DCGAN) is proposed to learn the data distribution from the given dataset and generate new data derived from the same distribution. Music can also be considered as a text of natural language requiring the network to learn the syntax of the sheet music completely and the dependencies among symbols\cite{garcia2019automatic}. Furthermore, as mentioned in \cite{de2022measuring}, the evaluations need a collection of metrics that can objectively describe structural properties of the music signal\cite{muller2011signal}. 

To generate music, this article introduces a novel AI-based music generator that combines LSTM networks and Fractional Fourier Transform (FrFT) techniques. The FrFT is a generalization of the traditional Fourier Transform that can handle non-stationary signals such as music signals\cite{boashash2018designing}. Unlike the traditional Fourier Transform, the FrFT provides a more flexible representation of signals, making it a useful tool for various signal processing tasks, including music generation. In particular, the FrFT has been shown to effectively capture the local structure of music signals, which is a critical factor in music generation. LSTMs, on the other hand, are a type of Recurrent Neural Network (RNN) that has been designed to handle sequential data\cite{sherstinsky2020fundamentals}. RNNs are well suited to processing sequential data, such as music signals, as they can maintain information about previous elements in the sequence over time. LSTMs have been applied in various fields including natural language processing, speech recognition, and music generation. The generator will be trained on a dataset of musical pieces, learning to generate new pieces that are similar in style to those in the training dataset. The FrFT will be used to provide a flexible representation of the music signals, while the LSTM network will be used to capture the local structure of the signals \cite{yao2018improved} and generate new pieces that are similar in style to the pieces in the training dataset.

One of the benefits of the proposed AI-based music generator is its ability to generate new pieces of music in a variety of styles, providing new inspiration for musicians and composers. The generator can also generate music in real-time, allowing for interactive music creation. Additionally, the generator has the potential to provide new possibilities for musical expression and creativity, enabling new forms of musical expression.

The implementation of the proposed music generator involves several steps, including pre-processing of the music signals, feature extraction using FrFT, and training of the LSTM network. The FrFT will be used to extract features from the music signals, representing them in a more flexible and informative form. The LSTM network will then be trained on the extracted features, learning to generate new pieces of music that are similar in style to those in the training dataset.

Certainly, here's a detailed breakdown of the key components discussed in the conclusion:

\textbullet Exploration of Time-Frequency Features and Signal Analysis This paper dives deep into the realm of music analysis, focusing on the intricate interplay of time and frequency components within musical signals. Through meticulous signal analysis and the extraction of time-frequency features, this research unveils the underlying intricacies and patterns inherent in music, shedding light on the fundamental essence of musical compositions.

\textbullet FrFT Feature Extraction: A significant aspect of this proposal involves the extraction of features using the FrFT. This technique offers a unique perspective on music representation, enabling the capture of intricate nuances and characteristics within musical data, thus enriching the music generation process.

\textbullet LSTM Music Generator: The core innovation of this paper lies in the creation of an AI-based music generator that integrates LSTM networks. This generator holds immense potential in revolutionizing music creation by leveraging deep learning to compose original pieces. It stands as a promising tool for musicians and composers seeking inspiration across diverse musical styles.

In Section 2, we introduce the music from the signal aspect and provide the FrFT method for time-frequency analysis. In addition, a LSTM formulation is shown for music prediction. In Section 3, a novel AI-music generator is proposed by combing the FrFT and LSTM techniques, including mining the deep features of the music from the fractional domains and using the LSTM networks for fitting the music. In section 4, we give the simulation results for the proposed scheme and show the loss values of the network training procedure. Simulation results concentrate that only small gaps are between the target music signal and predictions. Section 5 concludes the total paper.

\section{Music Signal and basic techniques}

\subsection{Music Signal}

Two popular file types used in the creation of music driven by AI are MIDI and WAV signals. While lacking audio samples, MIDI signals are a sort of digital transmission that encodes musical information such as note pitches and durations. Digital musical instruments and computer music apps frequently use it because it is a condensed and effective way to describe musical information.

MIDI signals are frequently used as input to neural networks or other types of machine learning algorithms in the process of music generation that is powered by AI. These algorithms can be taught to generate new MIDI signals by studying pre-existing musical compositions, which ultimately results in the creation of brand-new musical works that are unmatched in their originality. WAV signals, on the other hand, are a type of audio signal that contains sound waves that have been sampled. WAV signals, as opposed to MIDI signals, include actual audio samples and may be played back on digital audio devices. MIDI signals are not capable of this.

WAV signals can be used as the output of AI algorithms, which means they can be employed in music generation powered by AI. Using a software synthesizer, which transforms MIDI data into audio samples, it is possible to convert the MIDI signals that have been generated into WAV signals. This makes it possible for the music that was made to be played back as a sound wave and listened to as actual audio. In AI-driven music generation, MIDI and WAV signal each plays a significant role in the whole process. The artificial intelligence systems take in MIDI signals as inputs because they are condensed and effective representations of musical information. On the other hand, WAV signals make it possible for the user to hear the music that was generated by playing it back as actual audio. This is made possible by the fact that the music may be played again.

The conversion of musical information from one format to another is required to accomplish the translation between MIDI and WAV signals. The following procedures can be carried out to convert MIDI to WAV format: (1) Start by reading the MIDI signal, which includes information like the notes, durations, and velocities. (2) Convert the MIDI data to audio samples using a software synthesizer. This will result in a WAV signal that can be played back as a sound wave. The following procedures can be followed to convert WAV files to MIDI format: (1) Load the WAV signal, which is an audio signal that samples sound waves and can be found on your computer. (2) Create a MIDI signal by analyzing the WAV signal with a software tool or algorithm to extract musical information such as the pitches and durations of notes. This will result in the creation of the MIDI signal. The precise algorithms that are utilized for synthesizing and extracting musical information will depend on the specific requirements of the application, and their level of complexity and precision may vary depending on those requirements. Despite this, the fundamental procedures of reading the signals and converting from one format to another remain unchanged.

\subsection{FrFT}

Both the Fractional Fourier Transform (FrFT) and its inverse, the Inverse Fractional Fourier Transform (IFrFT), are mathematical tools that are used to examine signals in the time domain as well as the frequency domain, respectively\cite{sejdic2011fractional}. They are generalizations of the ordinary Fourier transform and its inverse, and they provide additional versatility when conducting signal analysis.

The FrFT of a function $f(t)$ is defined as\cite{faragallah2020investigation}:

\begin{equation}
F(\omega) = \frac{1}{\sqrt{2\pi}} \int_{-\infty}^{\infty} f(t) e^{-i\omega t} e^{-\frac{\pi}{2}i\text{sgn}(\omega) \alpha} dt
\end{equation}
where $\omega$ is the angular frequency, $\text{sgn}(\omega)$ is the sign function, and $\alpha$ is a parameter that defines the fractional order of the transform, with $-1 \leq \alpha \leq 1$.

Because it enables the rotation of the time and frequency axes inside the time-frequency plane, the Fourier transform (FrFT) makes it possible to do signal analysis in a variety of different domains. Analysis of signals that are non-stationary, have time-frequency localization or have a time-varying frequency can be accomplished with the help of the FrFT\cite{tao2006research}.

The IFFT of a function $F(\omega)$ is defined as:

\begin{equation}
f(t) = \frac{1}{\sqrt{2\pi}} \int_{-\infty}^{\infty} F(\omega) e^{i\omega t} e^{\frac{\pi}{2}i\text{sgn}(\omega) \alpha} d\omega
\end{equation}
and it provides a way to transform signals from the fractional Fourier domain back to the time domain. The IFFT can also be expressed in terms of the fractional Fourier series representation, as follows:

\begin{equation}
f(t) = \sum_{n=-\infty}^{\infty} c_n e^{-i\pi n \alpha} e^{i 2\pi n \frac{t}{T}}
\end{equation}
where $c_n$ is the Fourier coefficient and $T$ is the period of the signal.

The direct technique, the chirp-z transform method, and the FFT-based method are three examples of the many ways in which the FrFT and IFFT can be implemented. Each approach has a unique set of benefits and drawbacks, and selecting one approach is contingent upon the requirements of the particular application in question.

In conclusion, the Fractional Fourier Transform and its inverse offer an extremely useful tool for the investigation and processing of signals. They offer a greater degree of flexibility in the analysis of signals in various domains by permitting the time and frequency axes to be rotated. Both the fast Fourier transform (FrFT) and the fast Fourier transform (IFFT) are useful tools for signal processing and analysis because they may be used to analyze non-stationary signals, detect specific features, and transform signals between domains.

\subsection{LSTM}

Long Short-Term Memory (LSTM) networks are a type of Recurrent Neural Network (RNN) that are especially well-suited for processing sequences of data, such as time-series data or sequences of characters in the text. LSTM networks can also be used to store information for longer periods of time than traditional RNNs. In order to overcome the shortcomings of classic RNNs, which were unable to analyze long-term dependencies in sequential data in an efficient manner, LSTMs were designed as an alternative\cite{karim2017lstm}.

A standard RNN takes as input a sequence of vectors, $x_1, x_2, ..., x_T$, and produces a corresponding sequence of hidden states, $h_1, h_2, ..., h_T$. The hidden states are calculated using the following equation:

$$h_t = f(W_{hh}h_{t-1} + W_{xh}x_t + b_h)$$
where $W_{hh}$ and $W_{xh}$ are the weight matrices that connect the hidden state from the previous time step to the current hidden state and the input to the current hidden state, respectively. $b_h$ is the bias term and $f$ is the activation function, typically a non-linear function such as the sigmoid or tanh functions.

The output of the RNN at each time step is calculated using the following equation:

$$y_t = g(W_{hy}h_t + b_y)$$
where $W_{hy}$ is the weight matrix that connects the hidden state to the output, $b_y$ is the bias term, and $g$ is the activation function, typically a non-linear function such as the softmax function.

Traditional RNNs are able to handle sequences of data, but they struggle when it comes to dealing with dependencies in the data that are present over longer periods. This is a result of the fact that the hidden state is only updated at every time step and does not retain any information from earlier time steps. As a consequence of this, standard RNNs are frequently unable to process sequences of data adequately, particularly when the relationships between components in the sequence are separated in time\cite{subakan2021attention}.

This constraint is circumvented by LSTMs, which do so by integrating memory cells into the network. Input, forget, and output gates give the user the ability to exercise control over the memory cells, which are programmed to remember information from earlier time steps. It is possible to write information into the memory cell using the input gate, information can be forgotten using the forget gate, and information can be retrieved out of the memory cell using the output gate if those gates are properly configured.

The memory cells are updated using the following equation:
\begin{equation}
c_t = f_t \odot c_{t-1} + i_t \odot \tilde{c_t}
\end{equation}
where $c_t$ is the memory cell state at time step $t$, $f_t$ is the forget gate, $i_t$ is the input gate, and $\tilde{c_t}$ is the candidate memory cell state, calculated as:

\begin{equation}
\tilde{c_t} = tanh(W_{hc}h_{t-1} + W_{xc}x_t + b_c)
\end{equation}

The input, forget, and output gates are calculated using the following equations:

\begin{equation}
i_t = \sigma(W_{hi}h_{t-1} + W_{xi}x_t + b_i)
\end{equation}

\begin{equation}
    f_t = \sigma(W_{hf}h_{t-1} + W_{xf}x_t + b_f)
\end{equation}

\begin{equation}
o_t = \sigma(W_{ho}h_{t-1} + W_{xo}x_t + b_o)
\end{equation}
where $W_{hi}$, $W_{hf}$, $W_{ho}$, $W_{xi}$, $W_{xf}$, $W_{xo}$, $b_i$, $b_f$, and $b_o$ are the weight matrices and bias terms that connect the hidden state and input to the input gate, forget gate, and output gate, respectively. $\sigma$ is the sigmoid function, which is used to calculate the gate activations.

The hidden state is then calculated using the following equation:

\begin{equation}
h_t = o_t \odot tanh(c_t)
\end{equation}
where $\odot$ represents the element-wise multiplication. The output of the LSTM network is calculated using the same equation as for a traditional RNN:

\begin{equation}
y_t = g(W_{hy}h_t + b_y)
\end{equation}

LSTMs have found use in an extremely diverse set of applications, some examples of which are natural language processing, speech recognition, and the production of music, amongst many more. It has been demonstrated that they are especially adept at processing sequential data in which the relationships between components in the sequence are remote in time.

In the field of music generation, LSTMs have been used to generate music in many different forms, including pop, jazz, and classical. The network is taught to generate new pieces of music that are similar in style to the pieces that are included in the training dataset by being exposed to a huge dataset containing various musical works\cite{minu2022lstm}. When it comes to processing sequential data, including music, LSTMs are an extremely useful tool. They have been put to use in a wide variety of applications, such as the generation and transformation of music, and it has been demonstrated that they are particularly successful when it comes to dealing with long-term dependencies in sequential data.

\section{Scheme of the proposed AI-Music generator}

\subsection{Deep mining of music features based on FrFT}

\begin{figure}[h]
\centering
\includegraphics[width=6cm]{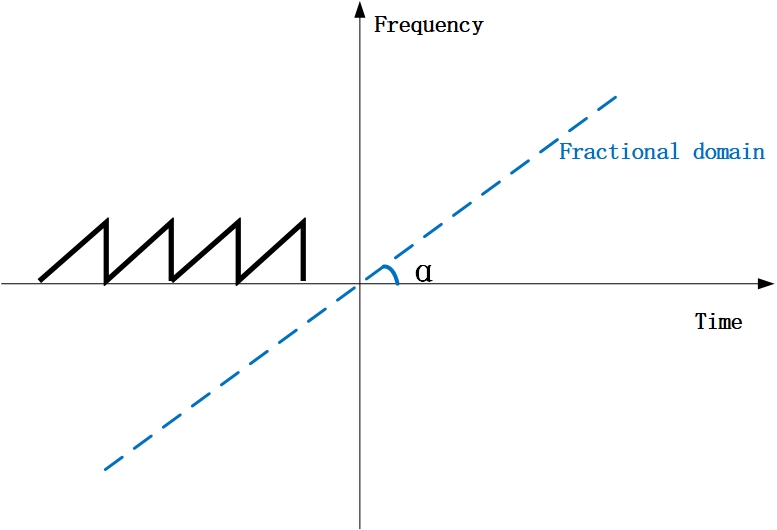}
\caption{Fractional domain and angle $\alpha$}\label{Fractional}
\end{figure}
Because the fundamental components of music are sound signals operating at a variety of frequencies, the properties of pieces of music that are played for a specific amount of time can be stated correspondingly in the time domain and the frequency domain. Additionally, some properties are unique to both the time domain and the frequency domain. As a result, doing extensive research into the properties of the time and frequency domains of music signals has the potential to enhance the overall performance of a musical composition.

The ideas behind the time domain and the frequency domain have been expanded as a result of the development of fractional Fourier changes. We can draw the time-frequency domain plane, also known as the fractional Fourier domain, by assigning time in the time domain to the X-axis and frequency in the frequency domain to the Y-axis of the time-frequency domain plane. 
Its important parameter is the fractional order domain between the time axis and the frequency axis, with the important Angle parameter $\alpha$, as shown in the figure \ref{Fractional}.

\begin{figure}[h]
\centering
\subfigure[Time-domain]{%
    \includegraphics[width=\columnwidth]{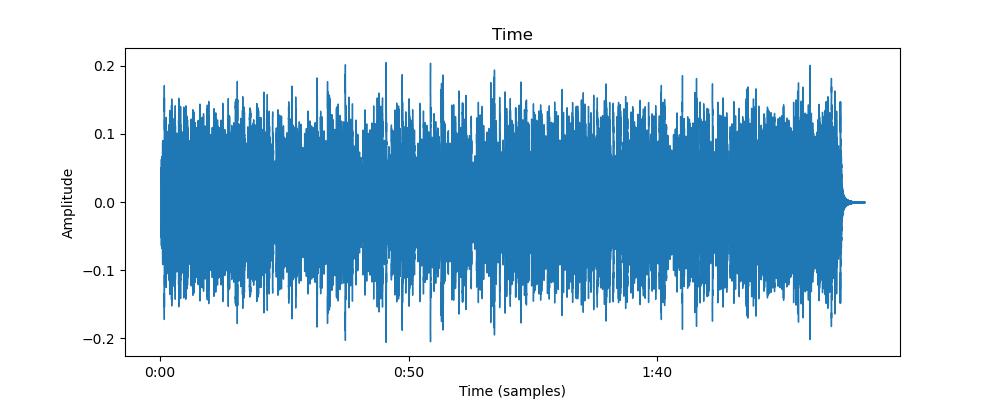}%
}
\subfigure[STFT]{%
    \includegraphics[width=\columnwidth]{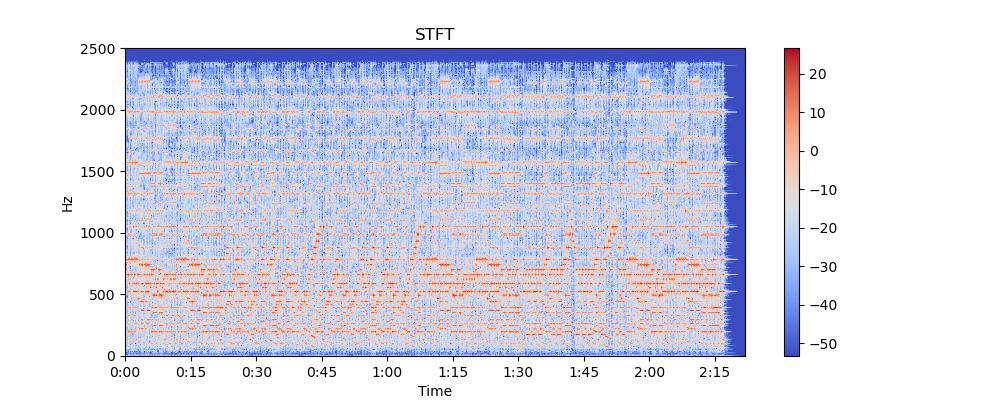}%
}
\subfigure[FrFT-real]{%
    \includegraphics[width=\columnwidth]{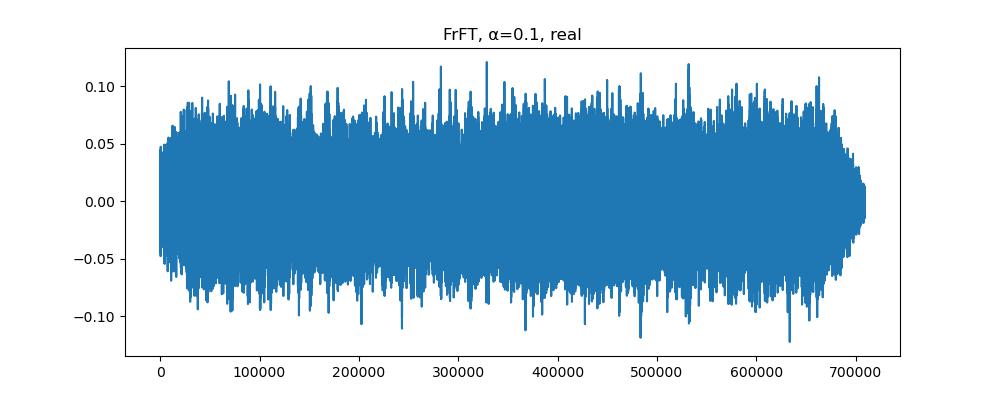}%
}
\subfigure[FrFT-imag]{%
    \includegraphics[width=\columnwidth]{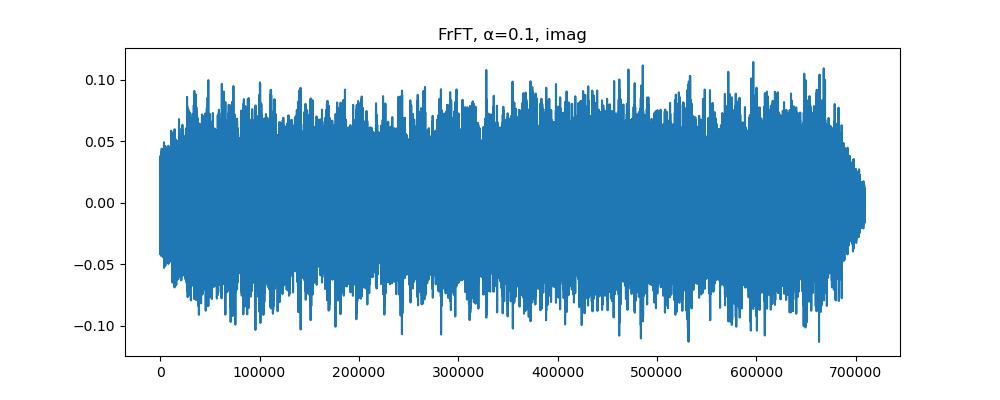}%
}
\caption{Feature maps of Clifford Adams' music sample.}
\label{fig:clifford}
\end{figure}

\begin{figure}[htbp]      % Give a unique label
% \quad

% \quad
\subfigure[Time]{\includegraphics[width=\columnwidth]{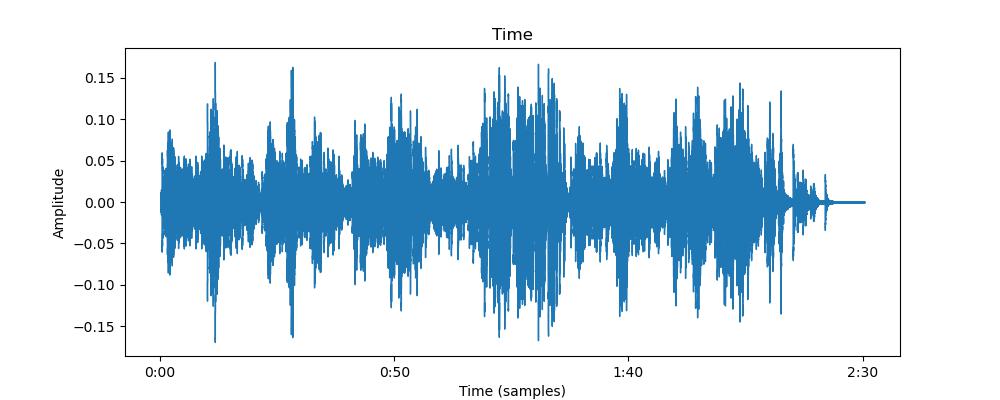}
}

\subfigure[STFT]{\includegraphics[width=\columnwidth]{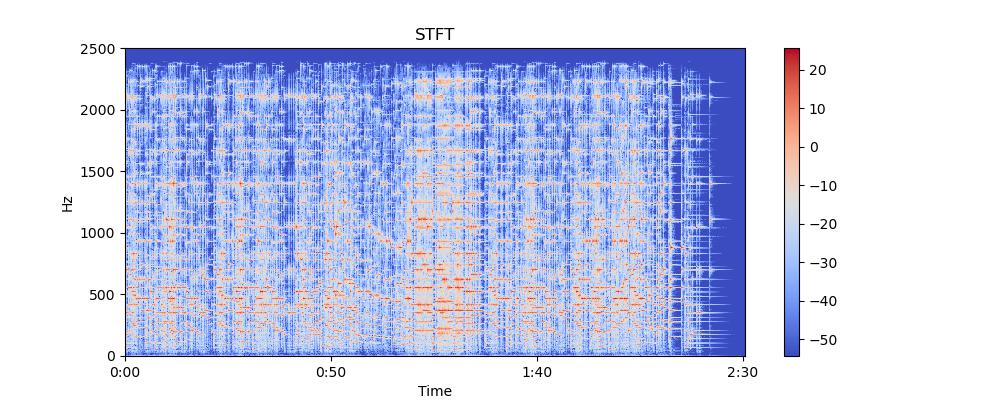}
}

\subfigure[FrFT-real]{\includegraphics[width=\columnwidth]{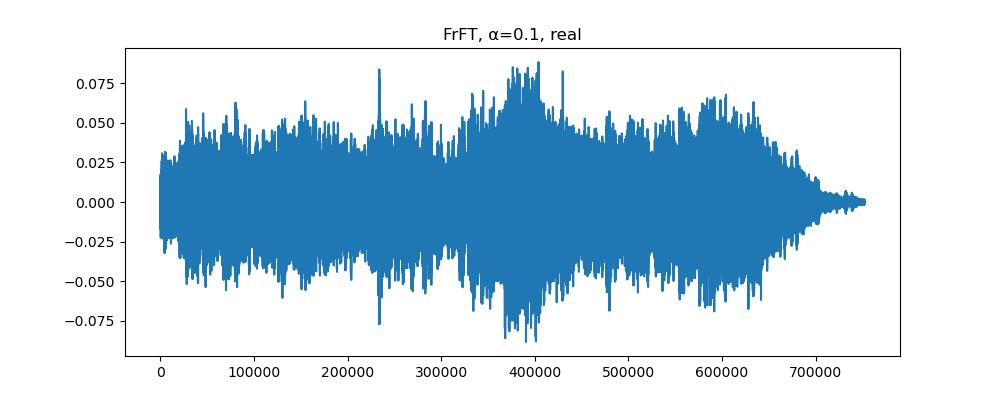}
}

\subfigure[FrFT-imag]{\includegraphics[width=\columnwidth]{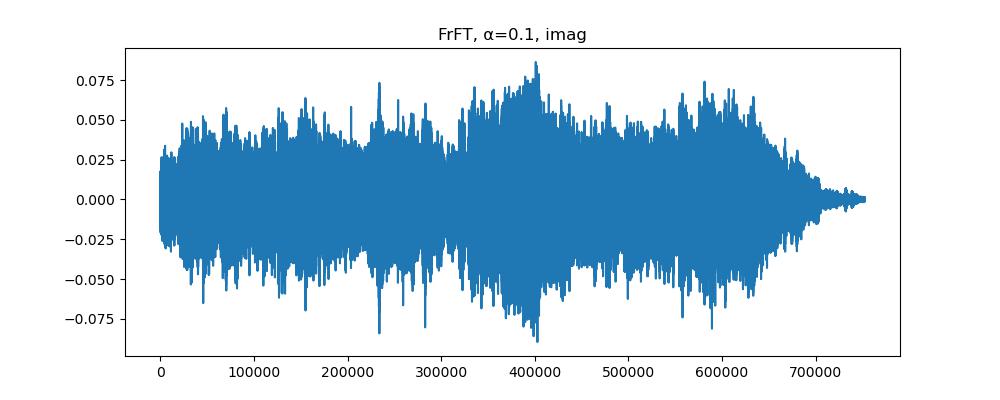}
}
\caption{Music mining demo: Evening Invocation\label{Demo} }
\end{figure} 

Here are the results of a deep dig into two piano pieces, Hanky Pank and Evening Invocation. Figure \ref{Demo}(a) and \ref{Demo}(b) are the time domain waveform, Figure \ref{Demo}(c) and \ref{Demo}(d) are the time-frequency domain result processed by STFT, Figure \ref{Demo}(e) and \ref{Demo}(f) are the real part of the result processed by FrFT. Figure \ref{Demo}(g) and \ref{Demo}(h) are the imaginary parts of the result after FrFT processing. We note that different music has different time-domain characteristics and different frequency-domain characteristics, which can be reflected in the above results. Therefore, the above processing results can be used in the subsequent training of LSTM network to predict the change rule of music, to realize the automatic music composing based on AI.

\subsection{The proposed scheme}

\begin{figure*}[h]
\centering
\includegraphics[width=15cm]{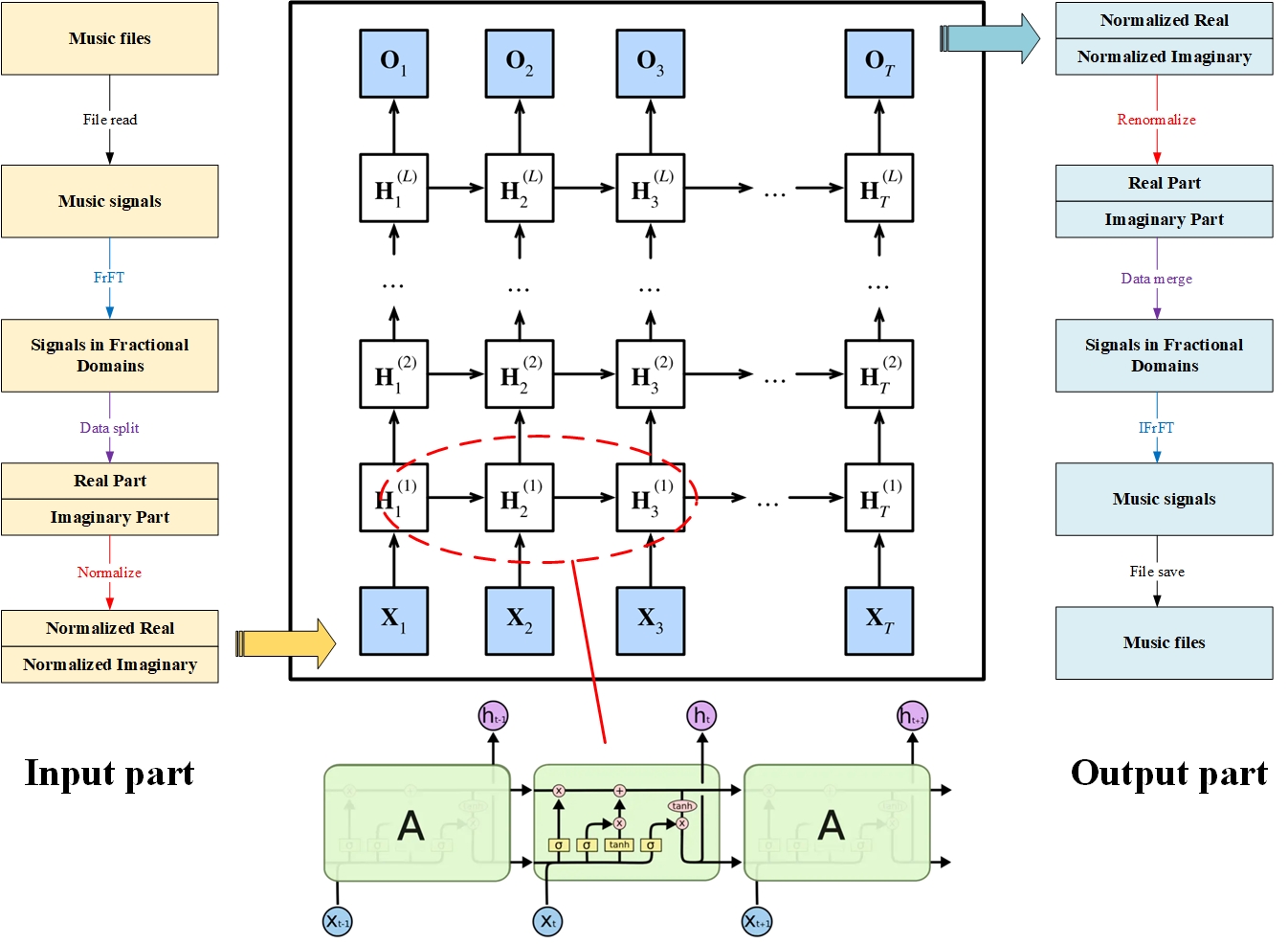}
\caption{Scheme of the proposed AI-Music generator}\label{Scheme}
\end{figure*}

The artificial intelligence-based music generator that has been proposed has been modeled as indicated in this section. The input part, the AI processing section, and the output part make up the three primary components that comprise the framework of the entire network. The processing of the music signal and the generation of data that is appropriate for the training of the lstm network are the primary goals of both the input and the output parts of the algorithm. For the purpose of AI-based music prediction, often known as music generation, the multi-layer LSTM network is utilized.

The selection of LSTM was motivated by its advantageous characteristics, such as its capacity to effectively process lengthy sequences, mitigate the issues of disappearing gradients and gradient explosion, and accommodate input across various temporal scales. The model demonstrates high proficiency in handling lengthy sequences, such as textual and auditory data. The gating mechanism effectively retains and selectively utilises long-term memory, while also accommodating information across various time intervals. Consequently, this model proves to be a robust tool for processing time series data.

As shown in Figure \ref{Scheme}, the input of the current time step as well as the hidden state of the previous time step are both sent as data into the gates of the long and short memory network. These data are then processed by three fully connected layers with sigmoid activation functions to calculate the values of the input gate, forgetting gate, and output gate, respectively. Each of the three gates has values that fall within the range of (0, 1). We will accept input from the intermediate layer of the multi-layer LSTM network that we are building by stacking many layers on top of one another. This will allow us to increase the overall performance of the training process. The input at the lower level as well as the concealed variable from the time before. Through the use of processed music data as input and output in the LSTM network as a whole, it is possible to implement AI-based music prediction using the capabilities of AI.

In the input part, the first input is the music file, after reading the music signal. Then fractional Fourier transform is used to convert the music signal into a fractional domain signal. Then, the obtained signal in the fractional domain is split into real and imaginary parts to obtain the real and imaginary parts of the signal, which are normalized respectively and used as the input of network training.

Obtaining the signal that is produced by the AI network falls under the purview of the output section. The inverse normalization processing is carried out on the output signal using the same parameters, and then the real part of the signal and the imaginary part of the signal are merged to generate the matching complex signal. The appropriate output music signal is then obtained through the use of the inverse fractional Fourier transform, which is then saved as a music file through the usage of the file.

 \subsection{Train and test procedure}

The LSTM network is a form of recurrent neural network (RNN) that is capable of capturing patterns in sequences of data. It is utilized in the process of creating new musical compositions. The network is "trained" using a series of musical pieces, each of which is represented as a string of notes in the training dataset. The purpose of the training procedure is to optimize the network's parameters in such a way that it will be able to produce new, cohesive pieces of music that are comparable to the pieces that are included in the training data.

The training procedure typically involves the following steps:

\begin{enumerate}
    \item Data preparation: The individual pieces of music that make up the training data are then subjected to preprocessing and format conversion before being fed into the network. Quantizing the notes into a set number of possible pitches and durations and dividing the parts into overlapping sequences of a given length may be required for this step.

    \item Model definition: The LSTM network is defined, which includes the number of layers that make up the network (N), the size of the hidden state vectors, and the activation functions that are utilized by the network.

    \item Model training: Using an optimization approach like stochastic gradient descent, the network is trained on the preprocessed musical pieces to learn how to recognize them. The loss function that is commonly used to evaluate the performance of the network is based on the difference that can be found between the musical pieces that have been generated and the pieces that correspond to them in the training data. This study adopts the Mean Squared Error (MSE) as the chosen loss function, which is mathematically represented by the following formula.
    \added{
    \begin{equation}
\operatorname{MSE}=(1 / n) * \sum\left(y_{-} i_{-}-\hat{y}_{-} i\right)^{\wedge} 2
\end{equation}
    }
    where $n$ is the sample size, ${y_{-} i_{-}}$ is the true label, and ${\hat{y}_{-} i}$ is the predicted value of the model.

    \item Model evaluation: A validation set of musical pieces that were not included in the training process is utilized to test the trained network's performance on the validation set. This enables a comparison to be made between the network and alternative models of music generation, which then enables the network to be fine-tuned based on the findings of the evaluation.

\end{enumerate}

The testing procedure for an LSTM network used for music generation typically involves the following steps:

\begin{enumerate}
\item Data preparation: The test data is preprocessed and transformed into the same format as the training data.

\item Model evaluation: Based on the test data, the performance of the trained network is assessed using a number of different metrics, including accuracy, precision, recall, and F1 score. These metrics are used to measure the quality of the musical pieces that are generated by the network as well as to compare the performance of the network to that of other models that make music.

\item Model tuning: Based on the results of the evaluation, the network can be fine-tuned by adjusting the parameters or architecture of the network to improve its performance on the test data.

\end{enumerate}

In conclusion, the technique for the training and testing of an LSTM network that is utilized for the generation of music involves a number of steps, such as the preparation of data, the definition of the model, training, assessment, and tuning. The purpose of the procedure is to optimize the network's parameters in such a way that it will be able to produce high-quality musical pieces that are comparable to the musical pieces that are included in the training data.

An ablation experiment was conducted to eliminate the segment of the original music that underwent processing via the fractional Fourier transform. The training of the LSTM-based neural network is exclusively conducted on the input music signal. In this study, we conduct a comparative analysis and evaluation of the aforementioned two techniques. Subsequently, we apply the resulting loss function values to the test set. The baseline approach exhibits a signal loss function value of 0.0351, while our method demonstrates a signal loss function value of 0.0155.

\section{Simulation results}

The artificial intelligence-based music synthesizer that was proposed earlier will now be simulated in this section. In every one of our simulations, we utilized the GiantMIDI-Piano dataset, which is a collection of MIDI files featuring solo piano music. It is a big collection of high-quality piano recordings that have been collected from a variety of sources including commercial sample libraries and publically accessible MIDI resources. The dataset is put to use in the process of training and assessing machine learning models for music generation and music analysis. GiantMIDI-Piano is an important resource for study in the field of music technology because of its extensive size as well as its high level of sound quality.

We first sampled the data set at an adoption rate of 5000 Hz and then processed it using the FrFT transform with the parameter $\alpha=0.05$. The real part and the imaginary part of the obtained complex signal are split and normalized respectively. We take the obtained real and imaginary part signals as a group of 200 sampling points to construct a training and test set suitable for the applied LSTM network structure. In the improved LSTM network, we set the size of the hidden layer as 256 and the number of network layers as 4. In the training, the learning rate is 0.0003, the batch size is 32, and the music synthesis is performed at 1 point each time. In the simulation process, we set the epoch of training as 30. 

\begin{figure}[h]
\centering
\includegraphics[width=7.5cm]{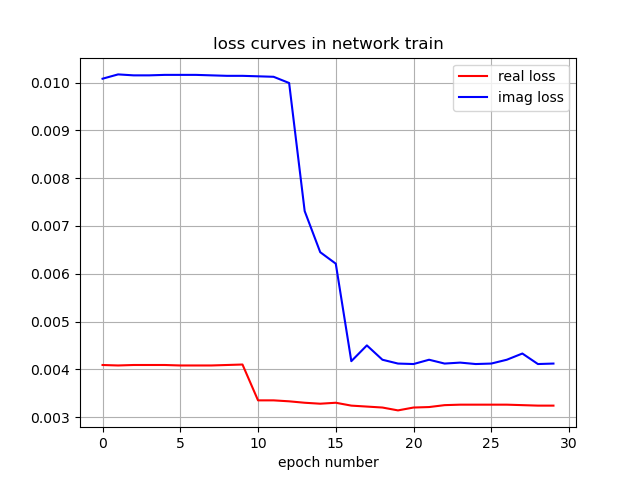}
\caption{Loss values in train procedure.}\label{Scheme007}
\end{figure}

\begin{figure}[h]
\centering
\includegraphics[width=7cm]{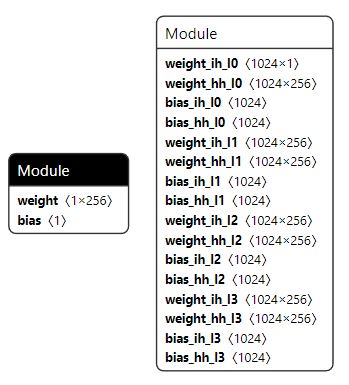}
\caption{The trained network structure.}\label{Net}
\end{figure}

We employed the proposed LSTM network for training and testing, and after first separating the training and test sets of the data set that was created by the music library, we used those sets to train and test the network. The image depicts the convergence curve that occurs throughout the training process. Included in this depiction is the convergence curve that occurs during the training of both the real part network and the imaginary part network. It can be seen that because FrFT exists, the real and imaginary parts of the produced features have different characteristics, and as a result, different loss function values and convergence rates. This is because the real and imaginary sections are obtained separately. In addition, the results of the training show that the final values of the loss function converge successfully, which demonstrates that the training of the network is successful.\added{The loss values in train procedure is shown in Figure \ref{Scheme007}.} The trained network structure is shown in Figure \ref{Net}.

\begin{figure}[htbp]      % Give a unique label
% \quad
\subfigure[train real]{
\includegraphics[width=7cm]{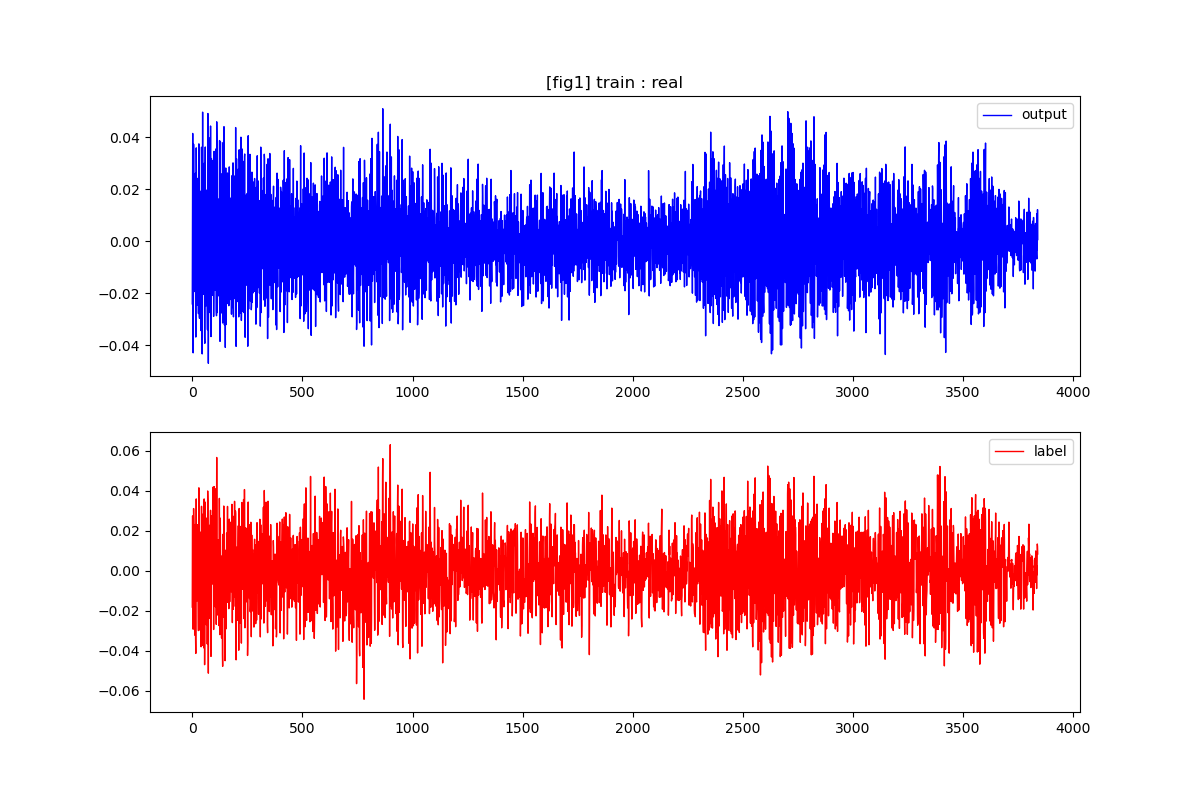}
}
% \quad
\subfigure[test real]{
\includegraphics[width=7cm]{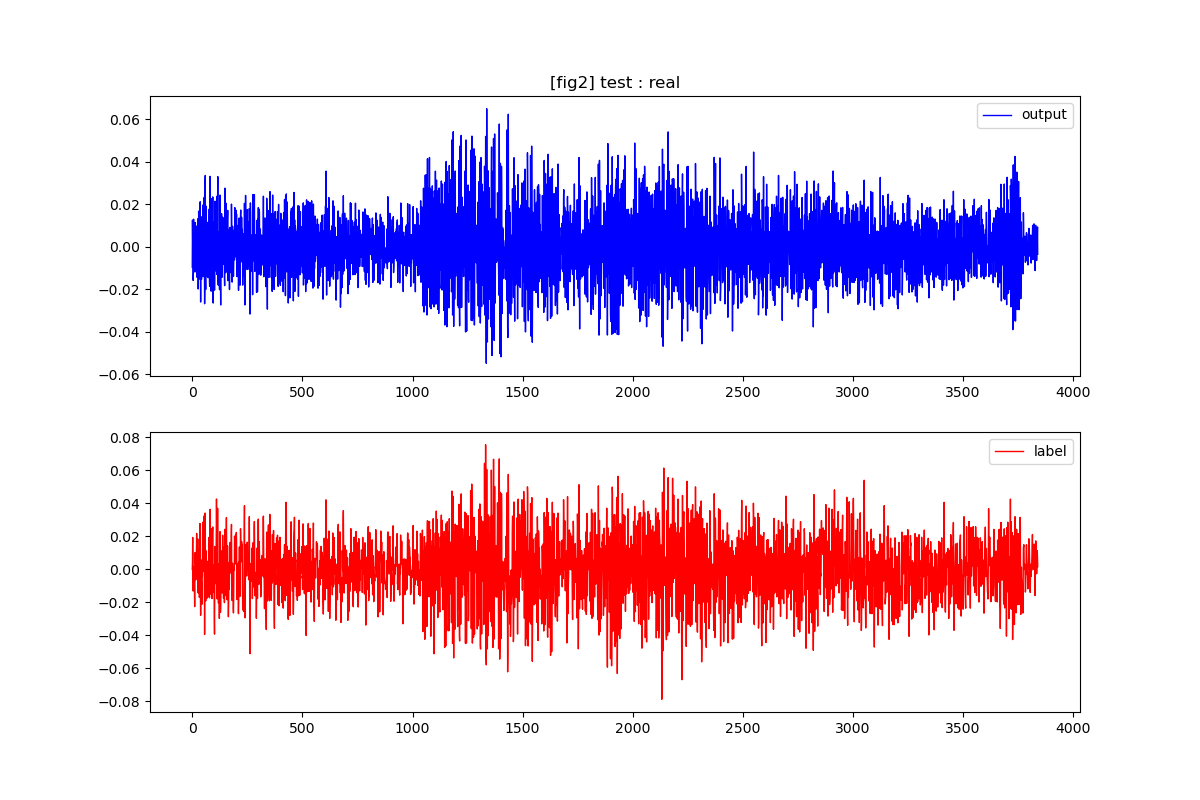}
}

\subfigure[train imaginary]{
\includegraphics[width=7cm]{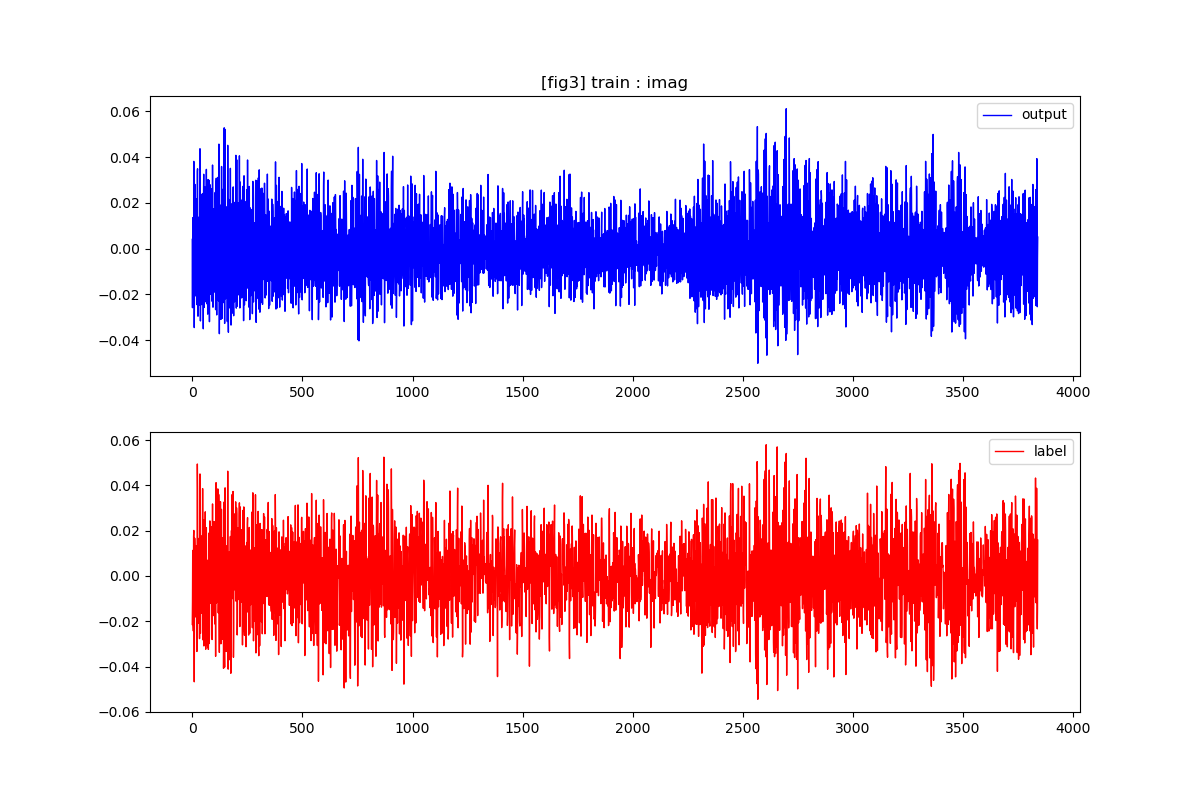}
}
\subfigure[test imaginary]{
\includegraphics[width=7cm]{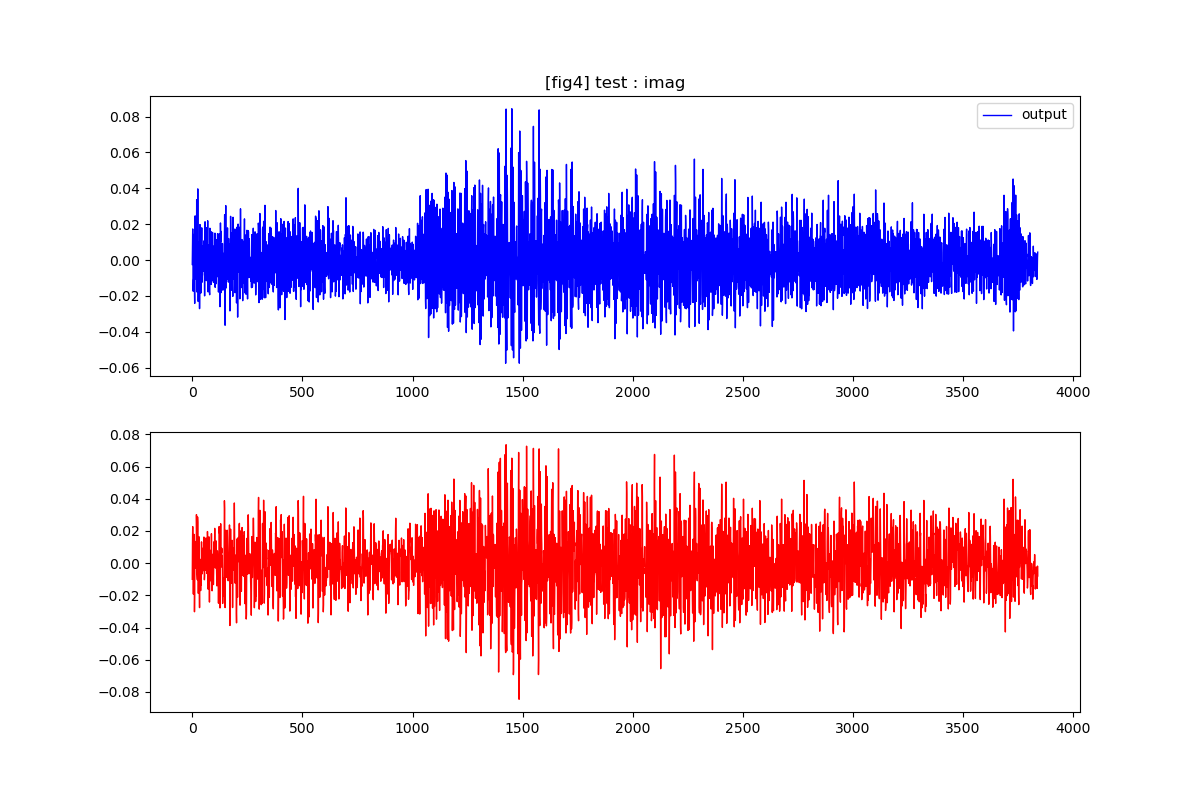}
}
\caption{Results of real and imaginary signals. \label{LSTM_realimag} }
\end{figure} 

To visualize the performance of music composition, we selected songs named 'Je t'aime Juliette' from the data set and analyzed their performance on the training and test sets. Figure \ref{LSTM_realimag} shows the results of training and testing LSTM networks with real and imaginary parts. We note that both the real and imaginary parts of the signals can fit the true value of the music, which shows the proposed LSTM can predict the real and imaginary parts well.

\begin{figure}[htbp]      % Give a unique label
% \quad
\subfigure[train real]{
\includegraphics[width=7cm]{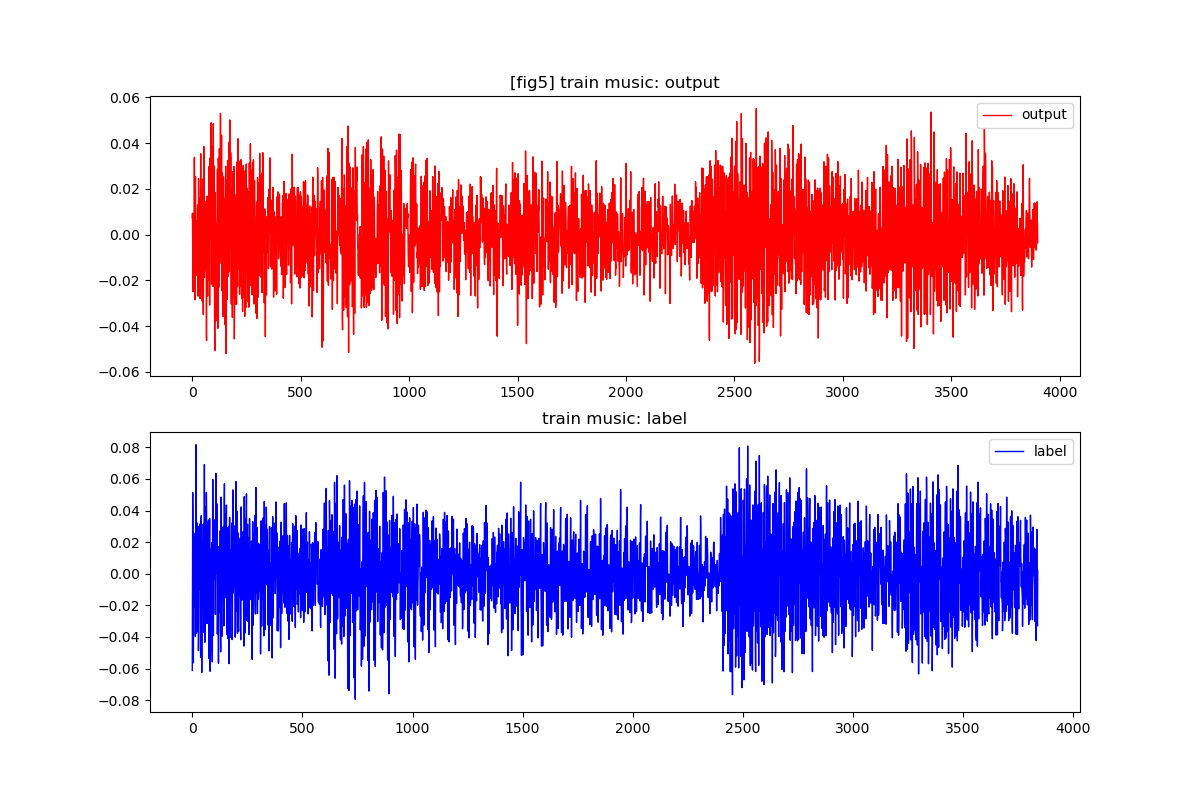}
}
% \quad
\subfigure[test real]{
\includegraphics[width=7cm]{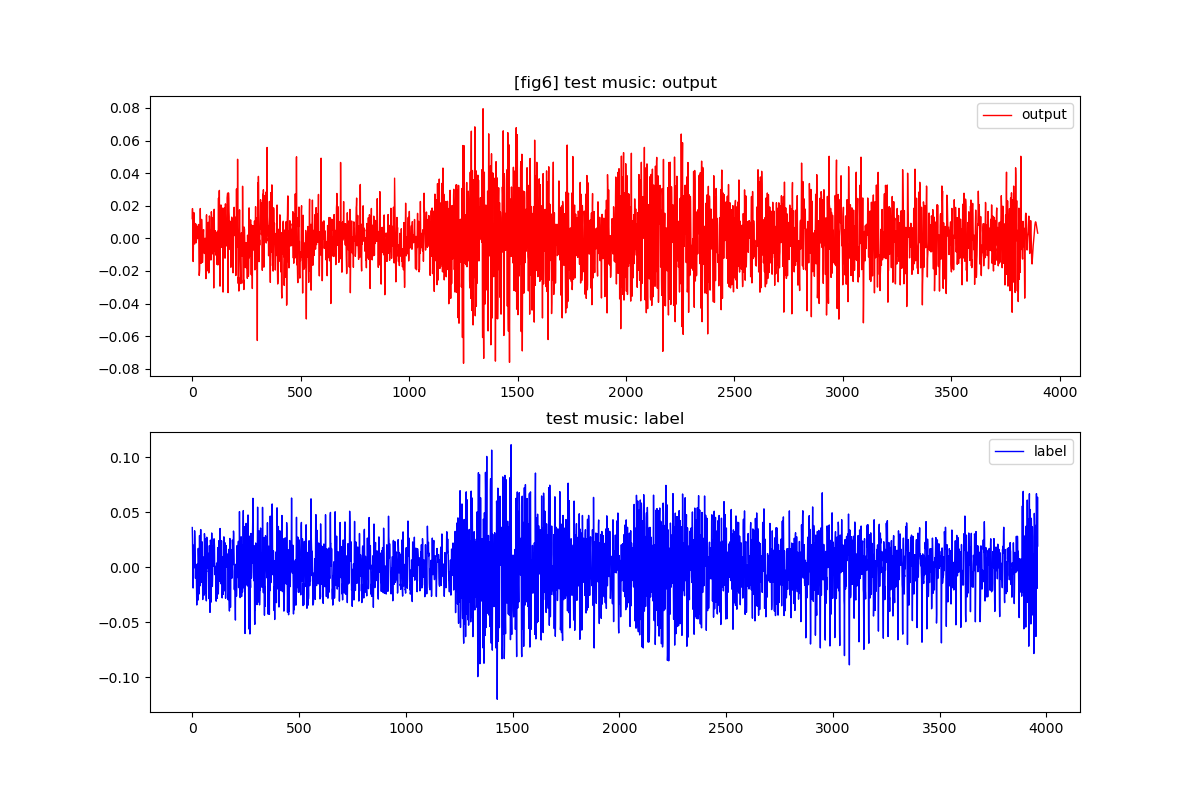}
}

\caption{Results of music signals. \label{LSTM_realimag} }
\end{figure} 

In the following stage, we test the suggested architecture's capability of being composed of other elements. The music waveforms from the training set and the test set are depicted side-by-side in the image. The signals that are produced as a result of the IFrFT operation are represented by the red curves, while the blue curves represent the original music signals before they were altered by the FrFt component. There is a high degree of resemblance between them, which suggests that the FrFT approach that was used is able to successfully realize the prediction of music, which allows it to successfully realize the function of automatic creation.

\section{Conclusion}

According to the findings of our research, utilizing a combination of FrFT and an LSTM network is a strategy that holds a lot of potential for the generation of music using AI. Our software can create music of a high quality that is very comparable to music that was created by humans. This system has the potential to be used in a variety of applications that are related to music, such as the writing of music, providing accompaniment, and the synthesis of music. The findings of this study indicate that there is significant potential for the development of AI-based music creation systems; however, additional research is required to perfect and perfect our system. The role of Music AI extends to fostering creativity, encouraging innovation, and expanding musical horizons. Additionally, it has a significant impact on music education by providing tailored exercises and demonstrations, facilitating students' skill development, and nurturing their creative potential.

\bmhead{Acknowledgments}

Not applicable.

\bmhead{Availability of data and materials}

Not applicable.

\bmhead{Funding}

This work was supported by Hainan Provincial Natural Science Foundation of China (Grant No. 723QN238).

\bmhead{Author information}
\bmhead{Authors and Affiliations}

Department of College of Music,Hainan Normal University, Longkunnan, Haikou, 570100, Hainan, China
(Li Ya, Yu Lei, Deng Xinyi)

Department of College of Foreign Languages,Hainan Normal University, Longkunnan, Haikou, 570100, Hainan, China
(Chen Wei)

Department of RD, Hainan Hairui Zhong Chuang Technol Co. Ltd., People, Haikou, 570228, Hainan, China
(Li Xiulai, Chen Chaofan)

\bmhead{Contributions}

LY, CW and LX have jointly participated in proposing the ideas, discussing the results, and writing and proofreading the manuscript. LY, CW and YL designed the core methodology of the study and carried out the implementation. DXY and CCF carried out the implementation of the algorithms and the experiments. All authors read and approved the final manuscript.

\bmhead{Competing interests}

The authors declare that they have no competing interests.

\bigskip
\


\end{document}